\documentclass[preprint2]{aastex}
\DeclareSymbolFont{msbm}{U}{msb}{m}{n} \DeclareMathSymbol{\C}{\mathalpha}{msbm}{'103}
\DeclareMathSymbol{\R}{\mathalpha}{msbm}{'122}
\DeclareMathSymbol{\Q}{\mathalpha}{msbm}{'121}
\DeclareMathSymbol{\Z}{\mathalpha}{msbm}{'132}
\DeclareMathSymbol{\N}{\mathalpha}{msbm}{'116}
\DeclareMathSymbol{\K}{\mathalpha}{msbm}{'113}

\newcommand{\myemailFF}{Fabio.Frescura@wits.ac.za}

\shorttitle{Pfaffian Structure with an Integrality Condition}
\shortauthors{Frescura and Lubczonok}

\begin{document}


\title{Pfaffian Structure with an Integrality Condition}



\author{F. A. M. Frescura}
\affil{Centre for Theoretical Physics, University of the Witwatersrand,
Private Bag 3, WITS 2050, South Africa} \email{\myemailFF}

\and

\author{G. Lubczonok}
\affil{Department of Mathematics (Pure and Applied), Rhodes University, P.O.
Box 94, Grahamstown 6140, South Africa.}




\begin{abstract}

Some Pfaffian manifolds admit the construction of an associated Weyl
line-bundle in which the lift $\overline{\alpha}$ of the Pfaffian structure
defines a 2-form $\overline{\omega}=d\overline{\alpha}$ which is basic. We
identify the conditions under which this construction is possible, implement
it, and investigate some properties of the foliated structure of these special
manifolds and of their canonical flows.

\end{abstract}


\keywords{Classical Mechanics --- Differential Geometry --- Contact Structure
--- Pfaffian Structure --- Hamiltonian Mechanics \\
\ \\ {\em PACS Numbers :}  02.40.-k, Ma; 45.20.Jj, -d
  }

\section{Introduction}
\label{section:Weyl}

In a Pfaffian structure, the 2-form $\omega=d\alpha$ is exact. However, in
general, it is not basic-exact. A 1-form $\beta$ is said to be {\em basic} if
$i_\xi \beta =0$ and ${\cal L}_\xi \beta =0$, where $\xi$ is the Liouville
vector field \citep{molino88,cp87}. We shall say that a two form $\phi$ is
{\em basic-exact} when it can be written as $\phi = d \beta$, where $\beta$ is
basic.

In certain special circumstances, a Pfaffian manifold permits the construction
of an associated Weyl bundle \citep{woodhouse80} in which the lift
$\overline{\alpha}$ of $\alpha$ defines a 2-form $\overline{\omega} = d
\overline{\alpha}$ which is basic-exact.  In this case, the construction of
the Weyl bundle is analogous to that of the standard one, and the success of
the construction relies on the existence of a particular type of \v{C}ech
cocyle.

In this paper, we implement this construction for this special case of
Pfaffian structure. In Section 2, we consider the concept of a \v{C}ech basic
2-cocycle for a Pfaffian Structure. In Section 3, we identify the conditions
under which a Pfaffian manifold admits the construction of an associated Weyl
bundle. The coditions include an integrality condition on the basic 2-cocycle.
We investigate the canonical objects admitted by this Weyl bundle and list
some of their properties.


\section{\v{C}ech Basic 2-cocycle of a Pfaffian Structure}

Let $\left\{U \right\}$ be a simple covering of $M$ by canonical charts, and
denote the coordinates of a canonical chart by $\{x^0, x^1, ..., x^n, p_1, ...
, p_n \}$. Then in each chart $U$ of this covering we have
\begin{eqnarray*}
  \alpha = dx^0 + p_i dx^i
\end{eqnarray*}
and
  \[ \omega = dp_i \wedge dx^i  \]
where, in accordance with the Einstein summation convention, a Latin index
repeated once in superscript position and once in subscript position indicates
a summation on the index range $\{1,2,..., n\}$. Since $\omega$ is exact, its
restriction to $U$ must possess a local potential $\beta_U$ such that on $U$
 \[ \omega = d\beta_U \]
In fact, we can always choose $\beta_U$ to be a locally basic 1-form by
putting
 \[ \beta_U = p_i dx^i  \]
Thus, on each $U$ of the covering, there exists at least one basic 1-form
$\beta_U$ such that
 \[ \omega = d\beta_U \]
First, consider all intersections $U\cap V$ of the charts of this covering. In
each $U\cap V \neq \emptyset$ we have
 \[ \omega = d \beta_U = d \beta_V  \]
so that
  \[ d(\beta_U - \beta_V) = 0  \]
Therefore there exists a function $\beta_{UV}$ on $U\cap V$ such that
 \[ \beta_U - \beta_V = d\beta_{UV}  \]
with each $\beta_{UV}$ being a basic function on $U\cap V$.

Now consider all threefold intersections $U\cap V \cap W$ of the charts of
this covering. In each $U\cap V \cap W \neq \emptyset$ we have
 \begin{eqnarray*}
  \beta_U - \beta_V = d\beta_{UV} \\
 \beta_V - \beta_W = d\beta_{VW} \\
 \beta_W - \beta_U = d\beta_{WU}
\end{eqnarray*}
so that
\[ d(\beta_{UV} + \beta_{VW} + \beta_{WU}) = 0  \]
and hence on each $U\cap V \cap W \neq \emptyset$ there exists a constant
function $c_{UVW}$ such that
\[\beta_{UV} + \beta_{VW} + \beta_{WU} = c_{UVW}  \]
The constant functions $c_{UVW}$ define a basic 2-cocycle in a basic \v{C}ech
cohomology \citep{woodhouse80, cp87}.


\section{The Integrality Condition}

In the general case there is little more that can be said about the existence
of the basic 2-cocycle described above.  However, in the special case when the
cocycle is cohomological to an integer basic cocycle, the manifold permits the
construction of an associated Weyl bundle.

Assume therefore that the basic 2-cocycle defined by the functions
$\{\beta_{UV}\}$ is cohomological to an integer basic cocycle.  We call this
assumption the {\em integrality condition}.  We can suppose without loss of
generality that the functions $\{\beta_{UV}\}$ have been chosen in such a way
that the constants $\{c_{UVW}\}$ are integers.  Put
\begin{eqnarray*}
 a_{UV} = e^{2\pi i\beta_{UV}}
\end{eqnarray*}
Then on each non-trivial $U \cap V \cap W$, the functions $a_{UV}$ satisfy the
relation
\[ a_{UV} a_{VW} a_{WU} = 1 \]
and thus define in a natural way a principal $U(1)$ bundle $P$ over $M$,
 \[ \pi : P \rightarrow M \]
 Since the form $\omega$ is exact, this bundle is trivial. However, it is not
necessarily {\em basic-trivial}.

The local basic potentials $\beta_U$ on $M$ define on $P$ a global 1-form
$\overline{\beta}$ given by
\[ \overline{\beta}|_U = ds + \pi^* \beta_U  \]
 where $\pi^* \beta_U$ is the pullback by the projection $\pi$ of $\beta_U$, $s$ is
the fibre parameter along the $S^1$ fibres of $P$, and $ds$ is its
differential. We then have
 \[ d \overline{\beta} = \overline{\omega} \]
where
 \[  \overline{\omega} = \pi^* \omega  \]
Put
\begin{eqnarray*}
 \overline{\alpha} &=& \pi^* \alpha \\
E &=& \frac{\partial}{\partial s}
\end{eqnarray*}
Thus $E$ is the vector field which defines the free action of the group $U(1)$
on $P$. Clearly $\overline{\beta}$ is a connection 1-form on $P$, with
curvature $\overline{\omega}$.

Denote by $X$ the horizontal lift of $\xi$. We thus have the following objects
on $P$: $\overline{\beta}$, $\overline{\alpha}$, $\overline{\omega}$; $E$,
$X$. In canonical coordinates $\{s, x^0, x^i, p_i \}$ on $\pi^{-1} (U)$ for
each $U\subset M$, we have
\begin{eqnarray}
 \overline{\beta} &=& d s + \overline{\beta}_U
 \label{2.1}
\end{eqnarray}
where
 \begin{eqnarray*}
 \lefteqn{  \overline{\beta}_U =  A^i(x^1,...,x^n,p_1,...,p_n)\ dp_i }\\
 &&\ \ \ \ + B_i(x^1,...,x^n,p_1,...,p_n)\ dx^i
 \end{eqnarray*}
 and
 \begin{eqnarray*}
 d \overline{\beta}_U &=& \overline{\omega}
 \end{eqnarray*}
Also
 \begin{eqnarray}
 \overline{\alpha} &=& d x^0 + p_i\ dx^i
 \label{2.2} \\
 \overline{\omega} &=& dp_i \wedge dx^i
 \label{2.3} \\
  E &=& \frac{\partial}{\partial s}
 \label{2.4} \\
  X &=& \frac{\partial}{\partial x^0}
 \label{2.4a}
\end{eqnarray}
 The quantities $\overline{\beta}$, $\overline{\omega}$, and $\overline{\alpha}$
 generate a closed, nowhere vanishing 1-form
 \begin{eqnarray}
 \overline{\theta} &=& \overline{\alpha}- \overline{\beta}
 \label{2.5}
\end{eqnarray}
and a pre-symplectic form
 \begin{eqnarray}
 \overline{\Omega} &=& \overline{\alpha} \wedge \overline{\beta} + \overline{\omega}
 \label{2.6}
 \end{eqnarray}
 with
 \begin{eqnarray}
 d \overline{\Omega} &=& \overline{\theta} \wedge \overline{\omega}
 \label{2.7}
 \end{eqnarray}
 Denote by $\sharp_{\overline{\Omega}}$ the raising operator for $
 \overline{\Omega}$. Then,
 \begin{eqnarray*}
 E= \sharp_{\overline{\Omega}}\ \overline{\alpha}
 \end{eqnarray*}
and
\begin{eqnarray*}
 X = \sharp_{\overline{\Omega}}\ \overline{\beta}
 \end{eqnarray*}
These quantities therefore do not generate a new vector field on $P$
independent of $E$ and $X$.

Using the representation of $\overline{\beta}$, $\overline{\omega}$,
$\overline{\alpha}$, $E$ and $X$ in canonical coordinates, we easily derive
the following relations:
\begin{enumerate}

\item Properties of $E$,
\begin{equation}
\label{6-4} \left. \begin{array}{cccccccccc}
 E \rfloor \overline{\alpha} &=& 0 & \ &  {\cal L}_E \overline{\alpha} &=& 0 \\
 E \rfloor \overline{\beta}  &=& 1 & \ &  {\cal L}_E \overline{\beta} &=& 0 \\
 E \rfloor \overline{\theta} &=& -1 & \ & {\cal L}_E \overline{\theta} &=& 0 \\
 E \rfloor \overline{\omega} &=& 0  & \ & {\cal L}_E \overline{\omega} &=& 0\\
 E \rfloor \overline{\Omega} &=& - \alpha &\ & {\cal L}_E \overline{\Omega} &=& 0 \\
\ [ E ,X ] &=& 0 & \   && \   &
\end{array} \right\}
\end{equation}

\item Properties of $X$,

\begin{equation}
\label{6-5} \left. \begin{array}{ccccccccccc}
 X \rfloor \overline{\alpha} &=& 1 &\ & {\cal L}_X \overline{\alpha} &=& 0 \\
 X \rfloor \overline{\beta} &=& 0  &\ & {\cal L}_X \overline{\beta} &=& 0 \\
X \rfloor \overline{\omega} &=& 0  &\ & {\cal L}_X \overline{\omega} &=& 0 \\
 X \rfloor \overline{\theta} &=& 1 &\ & {\cal L}_X \overline{\theta} &=& 0 \\
 X \rfloor \overline{\Omega} &=& \beta &\ & {\cal L}_X \overline{\Omega} &=& 0
\end{array} \right\}
\end{equation}

\item Properties of $Y = X+E$,

\begin{equation}
\label{6-6} \left. \begin{array}{ccccccccccc}
 Y \rfloor \overline{\alpha} &=& 1 & \ & {\cal L}_Y \overline{\alpha} &=& 0 \\
 Y \rfloor \overline{\beta}  &=& 1 & \ & {\cal L}_Y \overline{\beta} &=& 0 \\
 Y \rfloor \overline{\omega} &=& 0 & \ & {\cal L}_Y \overline{\omega} &=& 0\\
 Y \rfloor \overline{\theta} &=& 0 & \ & {\cal L}_Y \overline{\theta} &=& 0 \\
 Y \rfloor \overline{\Omega} &=& \theta & \ & {\cal L}_Y \overline{\Omega} &=& 0
\end{array} \right\}
\end{equation}

\end{enumerate}

We now assume furthermore that the Liouville field $\xi$ is complete. Then $X$
is also complete.

{\bf The $U(1)\times \R$ action on $P$:} With this assumption, the fields $E$
and $X$ generate a $U(1)\times \R$ action on $P$. Since $E$ and $X$ are
everywhere linearly independent, the orbits of this action are $S^1 \times
\R$, that is, cylinders, or $T^2 = S^1 \times S^1$. Each orbit projects onto a
corresponding orbit of $\xi$ in $M$.


\section{Consequences of the Integrality Condition}

The 1-form $\overline{\theta}$ is closed and nowhere zero. It therefore
defines a foliation $\{ F \}$ on $P$. Let $F$ be a leaf of $\{ F \}$. The $F$
clearly is a Pfaffian manifold with $\overline{\alpha}_F = \overline{\alpha}
|_F$, and $\overline{\omega}_F = d \left( \overline{\alpha}|_F \right)$. The
characteristic field on $F$ is $Y_F = Y|_F$. For each $p \in F$, the tangent
space $T_p F$ to the leaf projects one-to-one and onto $T_{\pi(p)} M$.

Given an orbit $O$ of the fields $E$ and $X$, the restriction
$\overline{\theta}_O$ defines a 1-dimensional foliation on $O$ generated by
the restriction $Y_O$ of $Y$ to $O$.

We now apply Tischler's structural theorem \citep{tischler70}.
 \begin{description}
\item[Case 1:] $\overline{\theta}= df$ and the leaves of the foliation are given by
$f=$ constant. Furthermore, the leaves define a trivial fibration of $P$ over
$\R$, and the flow of $E$ maps leaves of the foliation $\{ F \}$ onto leaves.
$E$ therefore defines a flow on $\R$. Since the flow for $E$ is periodic, so
also is the induced flow on $\R$. But the only periodic flow on $\R$ is the
constant flow. So the flow of $E$ must keep the leaves of $\{ F \}$ fixed. But
this is impossible, because $ E \rfloor \overline{\theta} = -1$. We have
therefore proved the following proposition.
\begin{quote}
\begin{description}
\item[Proposition 1:] The foliation defined by $\overline{\theta}$ cannot be
simple, or $\overline{\theta}$ is not an exact 1-form.
\end{description}
\end{quote}

\item[Case 2:] The foliation $\{F\}$ defines a locally trivial fibration
$p:P\rightarrow S^1$ over $S'$, all $S'\rightarrow S^1$. Since $E \rfloor
\overline{\theta}=-1$ and $X \rfloor \overline{\theta}=1$, the fields $E$ and
$X$ induce flows on $S'$ with no stationary points. So, the flows of $E$ and
$X$ act transitively on the leaves of the foliation $\{ F\}$, and all its
leaves are diffeomorphic closed submanifolds in $P$.

Consider first an orbit $O$ generated by $E$ and $X$ (or, $E$ and $Y$). From
(\ref{2.1}) and (\ref{2.2}), the restrictions to $O$ of $\overline{\alpha}$
and $\overline{\beta}$ are given respectively by $\overline{\alpha}|_O = ds$
and $\overline{\beta}|_O=dx^0$. The 2-form
\begin{eqnarray*}
    d s \otimes dx^0 + dx^0 \otimes d s
\end{eqnarray*}
defines a flat Riemannian metric on $O$, and the flows of $E$, $X$ and $Y$
restricted to $O$ are isometries of this metric, with $E$ and $X$ orthogonal.
The orbits of $E$ are circles, so the orthogonal orbits of $X$ are circles in
the case of a toral $O$ and generating lines in the case of a cylinder. Since
$Y=E+X$, the orbits of $Y$ are closed if $O$ is a torus and helices if $O$ is
a cylinder. From the construction of the Weyl bundle, all orbits of $E$ have
period 1.

The intersection of the leaves of $\{F\}$ with $O$ are precisely the orbits of
$Y$ on $O$. Let $T$ be the basic period of the flow induced by $X$ on the base
space $S'$, and let $T_O$ be the period of the flow of $X$ on $O$. Then $T_O$
is a multiple of $T$. This proves the following proposition.
\begin{quote}
\begin{description}
\item[Proposition 2:] The periods of periodic trajectories of $X$ are multiples of $T$.
\end{description}
\end{quote}

\item[Case 3:] All the leaves of $\{F\}$ are dense in $P$. In this case, the
consideration of Case 2 apply to $O$, but the conclusion stated in Proposition
2 does not follow.

\end{description}

Consider now the leaves of $\{F\}$ in cases 2 and 3 above. The flow of $E$ has
period 1, and $T_O$ denotes the the basic period of the flow of $X$ on the
base space $S'$. So, in Case 2, according to Proposition 2, there must be some
positive integer $\ell$ such that $\ell T_O=1$. This means that each leaf $F$
intersects $P_x = \pi^{-1}(x)$, $x\in M$, $\ell$ times. We have thus proved
the following proposition.
\begin{quote}
\begin{description}
\item[Proposition 3:] In Case 2, each leaf $F$ is an $\ell$-fold covering of $M$.
\end{description}
\end{quote}
In Case 3, each leaf of $F$ is dense in $P$. Consider $F\cap P_x$ for a given
$U(1)$ orbit $P_x$ over $x\in M$. Suppose that, for some interval $(t_1,t_2)$
of $P_x$, $F\cap (t_1,t_2)= \emptyset$. Since $F$ projects locally one-to-one
onto $M$, this implies that an open set of $P$ is disjoint from $F$. This is a
contradiction. We have therefore proved the following proposition.
\begin{quote}
\begin{description}
\item[Proposition 4:] In Case 3, the intersection $P_x \cap F$ is a dense subset of $P_x$.
\end{description}
\end{quote}
These results has consequences for the flow of $\xi$. Since the orbits of $X$
intersect those of $E$, we arrive at the following conclusion from Proposition
2.
\begin{quote}
\begin{description}
\item[Proposition 5:] The periods of periodic trajectories of $\xi$ are multiples of $T$.
\end{description}
\end{quote}





\begin{thebibliography}{99}

\bibitem[\protect\citeauthoryear{Craivreanu \& Puta}{1987}]{cp87}
Craivreanu, M., and Puta, M. (1987).  Cohomology classes and foliated
manifolds, in  {\em Nonlinear Analysis}, edited by Rassias, Th. M.,  World
Scientific Publishing Company, Singapore.

\bibitem[\protect\citeauthoryear{Molino}{1988}]{molino88} Molino, P., 1988,
Riemannian Foliations. {\em Progress in Mathematics}, {\bf 73}, Birkhauser.

\bibitem[\protect\citeauthoryear{Tischler}{1970}]{tischler70} Tischler, D., 1970,
Topology {\bf 9}, p 153-154.

\bibitem[\protect\citeauthoryear{Woodhouse}{1980}]{woodhouse80}
Woodhouse, N., 1980, {\em Geometric Quantisation}, Oxford University Press.

\end{thebibliography}
\end{document}